\documentclass[structabstract]{aa}
\usepackage{txfonts}
\usepackage{graphicx}
\usepackage{subfigure}

\usepackage{natbib}
\usepackage{lineno}
\usepackage{lscape}
\newcommand\T{\rule{0pt}{2.6ex}}
\newcommand\B{\rule[-1.2ex]{0pt}{0pt}}

\title{\emph{Fermi}-LAT constraints on the Pulsar Wind Nebula nature of \\HESS J1857$+$026}
\subtitle{}

\author{
R.~Rousseau$^{(1,2)}$ \and
M.-H.~Grondin$^{(3,4)}$ \and
A.~Van~Etten$^{(5)}$ \and
M.~Lemoine-Goumard$^{(1,2)}$ \and
S.~Bogdanov$^{(6)}$ \and
J.W.T.~Hessels$^{(7,8)}$ \and   
V.~M.~Kaspi$^{(9)}$ \and
Z.~Arzoumanian$^{(10)}$ \and
F.~Camilo$^{(6)}$ \and 
J.~M.~Casandjian$^{(11)}$ \and 
C.~M.~Espinoza$^{(12)}$ \and 
S.~Johnston$^{(13)}$ \and  
A.~G.~Lyne$^{(12)}$ \and 
D.~A.~Smith$^{(1)}$ \and 
B.~W.~Stappers$^{(12)}$ \and
G.~A.~Caliandro$^{(14)}$ 
}

\authorrunning{LAT collaboration}

\institute{
\inst{1}~ Centre d'\'Etudes Nucl\'eaires de Bordeaux Gradignan, Universit\'e Bordeaux 1, CNRS/IN2p3,  33175, Gradignan, France\\ 
\inst{2}~Funded by contract ERC-StG-259391 from the European Community\\ 
\inst{3}~Max-Planck-Institut f\"ur Kernphysik, D-69029 Heidelberg, Germany\\ 
\inst{4}~Landessternwarte, Universit\"at Heidelberg, K\"onigstuhl, D 69117 Heidelberg, Germany\\ 
\inst{5}~W. W. Hansen Experimental Physics Laboratory, Kavli Institute for Particle Astrophysics and Cosmology, Department of Physics and SLAC National Accelerator Laboratory, Stanford University, Stanford, CA 94305, USA\\ 
\inst{6}~Columbia Astrophysics Laboratory, Columbia University, New York, NY 10027, USA\\ 
\inst{7}~ASTRON, the Netherlands Institute for Radio Astronomy, Postbus 2, 7990 AA, Dwingeloo, The Netherlands\\ 
\inst{8}~Astronomical Institute ``Anton Pannekoek", University of Amsterdam, Science Park 904, 1098 XH Amsterdam, The Netherlands\\  
\inst{9}~Department of Physics, McGill University, Montreal, PQ, Canada H3A 2T8\\ 
\inst{10}X-ray Astrophysics Laboratory and CRESST, NASA Goddard Space Flight Center, Code 662, Greenbelt, MD 20771, USA \\
\inst{11}~Laboratoire AIM, CEA-IRFU/CNRS/Universit\'e Paris Diderot, Service d'Astrophysique, CEA Saclay, 91191 Gif sur Yvette, France\\ 
\inst{12}~Jodrell Bank Centre for Astrophysics, School of Physics and Astronomy, The University of Manchester, M13 9PL, UK\\ 
\inst{13}~CSIRO Astronomy and Space Science, Australia Telescope National Facility, Epping NSW 1710, Australia\\ 
\inst{14}~Institut de Ci\`encies de l'Espai (IEEE-CSIC), Campus UAB, 08193 Barcelona, Spain\\
\email{rousseau@cenbg.in2p3.fr, lemoine@cenbg.in2p3.fr, Marie-Helene.Grondin@mpi-hd.mpg.de, ave@stanford.edu} \\
}

\begin{document}
\date{Received  /Accepted }


\abstract
{Since its launch, the \emph{Fermi} satellite has firmly identified 5 pulsar wind nebulae plus a large number of candidates, all powered by young and energetic pulsars. HESS J1857$+$026 is a spatially extended $\gamma$-ray source detected by H.E.S.S. and classified as a possible pulsar wind nebula candidate powered by PSR J1856$+$0245.}{We search for $\gamma$-ray pulsations from PSR J1856$+$0245 and explore the characteristics of its associated pulsar wind nebula.}{Using a rotational ephemeris obtained from the Lovell telescope at Jodrell Bank Observatory at $1.5$ GHz, we phase$-$fold 36 months of $\gamma$-ray data acquired by the Large Area Telescope (LAT) aboard \emph{Fermi}. We also perform a complete $\gamma-$ray spectral and morphological analysis.}{No $\gamma$-ray pulsations were detected from PSR J1856$+$0245. However, significant emission is detected at a position coincident with the TeV source HESS J1857$+$026. The $\gamma$-ray spectrum is well described by a simple power-law with a spectral index of $\Gamma = 1.53 \pm 0.11_{\rm stat} \pm 0.55_{\rm syst}$ and an energy flux of $G(0.1$--$100$ GeV$)=(2.71 \pm 0.52_{\rm stat} \pm 1.51_{\rm syst}) \times 10^{-11}$ ergs cm$^{-2}$ s$^{-1}$. The $\gamma$-ray luminosity is $L_{PWN}^{\gamma} (0.1$--$100$ GeV$)=(2.5 \pm 0.5_{stat} \pm 1.5_{syst}) \times 10^{35} \left( \frac{d}{9 kpc} \right)^2$ ergs s$^{-1}$, assuming a distance of 9~kpc. This implies a $\gamma-$ray efficiency of $\sim$ 5 $\%$ for $\dot{E}=4.6 \times 10^{36}$ erg $s^{-1}$, in the range expected for pulsar wind nebulae. Detailed multi-wavelength modeling provides new constraints on its pulsar wind nebula nature.}{}
\authorrunning{Rousseau et al.}
\titlerunning{\emph{Fermi}-LAT constraints on the PWN nature of HESS J1857+026}
\maketitle
\section{Introduction}
Pulsar wind nebulae (PWNe) are bubbles of shocked relativistic particles produced by the interaction of the pulsar's wind with the surrounding medium \citep{Gaensler 2006}. Since 2003, the continuous observations of the Galactic Plane by $\breve{C}$erenkov telescopes have yielded the detection of more than 60 Galactic TeV sources. Among them, PWNe are the dominant class with 29 firm identifications. In the GeV energy range, 7 PWNe have been firmly identified by the \emph{Fermi}-LAT, all of them having their Inverse Compton (IC) peak at energies higher than 100~GeV. They are all powered by energetic pulsars and their $\gamma$-efficiencies are $\sim 1\%$, consistent with TeV observations \citep{Ackermann 2011}.

The presence of a pulsar close to the source position is an important clue to identify a PWN, which often requires information from the radio/X-ray wavelengths. Radio/X-ray PWNe are often associated with TeV extended sources offset from their pulsars, which can be explained by an inhomogeneous environment \citep{Hinton 2010}. In such sources, TeV radiation can be explained by IC scattering of accelerated leptons on ambient photon fields (CMB, IR, ...) or by $\pi ^0$ decay from the interaction of accelerated hadrons with nuclei of the interstellar medium.

HESS J1857+026 is a very high energy (VHE) $\gamma$-ray source detected by H.E.S.S. during the Galactic Plane Survey \citep{Aharonian 2008}. The extended ($\sim$ 0.11$\degr$) TeV source was identified as a PWN candidate after the discovery of PSR J1856+0245 (offset $\sim$ 0.12$\degr$) in the Arecibo PALFA survey \citep{Hessels 2008} with a Dispersion Measure (DM) of 622 cm$^{-3}$ pc. Recently, MAGIC reported a measured extension in the 0.2--1~TeV energy range significantly larger (0.22$\degr$) than the extension reported by H.E.S.S. in the 0.6--80~TeV energy range \citep{Klepser 2011}. 
PSR J1856+0245 is an energetic pulsar ($\dot{E}=4.6 \times 10^{36}$ erg $s^{-1}$) located in a crowded region, 1.3$\degr$ from the bright SNR W44 (Abdo et al. 2010) and 0.6$\degr$ from the fainter SNR HESS J1858+020 on which only an upper limit could be set using \emph{Fermi}-LAT data \citep{Torres 2011}. Significant emission coincident with HESS J1857+026 was observed above 100 GeV using \emph{Fermi}-LAT observations \citep{Neronov 2010}.

Here, we report in detail GeV observations of the HESS J1857+026/PSR J1856+0245 system using \emph{Fermi}-LAT observations and discuss their implications for the nature of the source.

\section{LAT description and data selection}
The LAT is a $\gamma$-ray telescope that detects photons by conversion into electron-positron pairs and operates in the energy range between 20 MeV and 300 GeV. Details of the instrument and data processing are given in \cite{Atwood 2009}. The on-orbit calibration is described in \cite{Abdo 2009 a}.

The following analysis was performed using 36 months of data collected from August 4, 2008 to August 31, 2011 within a $10 \times 10\degr$ square around the position of 
HESS J1857+026 aligned with Galactic coordinates. We excluded $\gamma$-rays coming from a zenith angle larger than 100$\degr$ because of possible contamination from secondary $\gamma$-rays from the Earth's atmosphere \citep{Abdo 2009 b}. We used the P7 V6 Instrument Response Functions (IRFs), and selected the `Source' events which correspond to the best compromise between the number of selected photons and the charged particle residual background for the study of point-like or slightly extended sources. 

\section{Data analysis}
\subsection{Timing analysis of PSR~J1856+0245}

With its large spin-down power, PSR~J1856+0245 is one of the more energetic radio pulsars known.  Its spin period of 80.9~ms and characteristic age of 20.6~kyr are similar to those of the Vela pulsar. The DM and NE2001 electron density model of the Galaxy assign PSR J1856+0245 a distance of $\sim$~9 kpc \citep{Cordes 2002}. 

This pulsar is not monitored as part of the LAT pulsar timing campaign \citep{Smith 2008}, as it was discovered subsequently, but has nevertheless been regularly observed with the Lovell telescope at Jodrell Bank Observatory \citep{Hobbs 2004}. 
The ephemeris of  PSR~J1856+0245 used in the analysis was obtained using 82 observations at 1.5~GHz made with the Lovell telescope between May 4, 2008 and September 4, 2011. The event arrival times were corrected to the Solar System Barycenter using the JPL DE405 Solar System ephemeris.
The $\mathtt{TEMPO2}$ timing package \citep{Hobbs 2006} was then used to build the timing solution. We fit the radio times of arrival (TOAs) to the pulsar rotation frequency and first four derivatives (in order to remove timing noise). By including a fourth derivative the RMS of the timing residuals decreases by about 30\%, from 1.55~ms  to  1.08~ms, equivalent to a decrease from 19 to 13 milliperiods. The reduced $\chi^2$ of the fit also decreases significantly, from 22 to 5. The fit results are summarized in Table \ref{tab1}. This timing solution will be made available through the \emph{Fermi} Science Support Center\footnote{FSSC:http://fermi.gsfc.nasa.gov/ssc/data/access/lat/ephems/} (FSSC).
\begin{table}[!t]
\begin{center}
\begin{tabular}{|l|l|}
  \hline
  Parameter & Value \\
  \hline
  $\nu$ (Hz) & $12.3597551142(1)$ \\
  $\nu^{(1)}$ (Hz s$^{-1}$) & $-9.48698(1) \times 10^{-12}$ \\
  $\nu^{(2)}$ (Hz s$^{-2}$) & $1.6585(9)\times 10^{-22}$\\
  $\nu^{(3)}$ (Hz s$^{-3}$) & $2.47(7) \times 10^{-30}$\\
  $\nu^{(4)}$ (Hz s$^{-4}$) & $1.32(7) \times 10^{-37}$\\
  \hline
  DM (cm$^{-3}$ pc) & 622\\
  Period epoch (MJD) & 55128\\
  Start time (MJD) & 54615\\
  End time (MJD) & 54570\\
  Number of TOAs & 82\\
  TOA rms (ms) & 1.08\\
  \hline
\end{tabular}
\end{center}
\caption{Parameters of the fit of the TOAs. $\nu$ correspond to the rotational frequency of the pulsar and $\nu^{(i)}$ its time derivative of order i. In parentheses are the 1 $\sigma$ uncertainty on the least-significant digits quoted, from $\mathtt{TEMPO2}$.}
\label{tab1}
\end{table}

For the LAT analysis, photons with energies above 100 MeV and within a radius of 1.0$^{\circ}$ of the radio pulsar position $\alpha$(J2000) = 18$^h$56$^m$50.937$^s$, $\delta$(J2000) = +02$\degr$45$^\prime$ 47.046$^{\prime \prime}$ were selected using an energy-dependent cone of radius $\displaystyle \theta <\max(5.12\degr \times (E/100 \, {\rm MeV})^{-0.8},0.2\degr)$  and phase-folded using the radio ephemeris previously described. This choice takes into account the instrument performance and improves the single-photon signal-to-noise ratio over a broad energy range. No significant pulsation was detected for all tested energy bands (100 MeV -- 300 GeV, 100 MeV -- 300 MeV, 300 MeV -- 1 GeV, $>$ 1 GeV). Following the procedure used by \cite{Romani11}, we fitted a point source at the position of PSR~J1856+0245 in the 0.1 -- 1 GeV energy range assuming a power-law of index 1.62 and a cut-off energy at 2.8 GeV to derive a 99$\%$ Bayesian upper limit on the flux of $~3.27 \times 10^{-8}$ ph cm$^{-2}$ s$^{-1}$ , well below typical $\gamma$-ray fluxes reported for pulsars detected by \emph{Fermi}-LAT \citep{Abdo 2010 c}.
\begin{figure}[h!]
\centering
\subfigure{
\includegraphics[width=0.36\textwidth, height=0.356\textwidth]{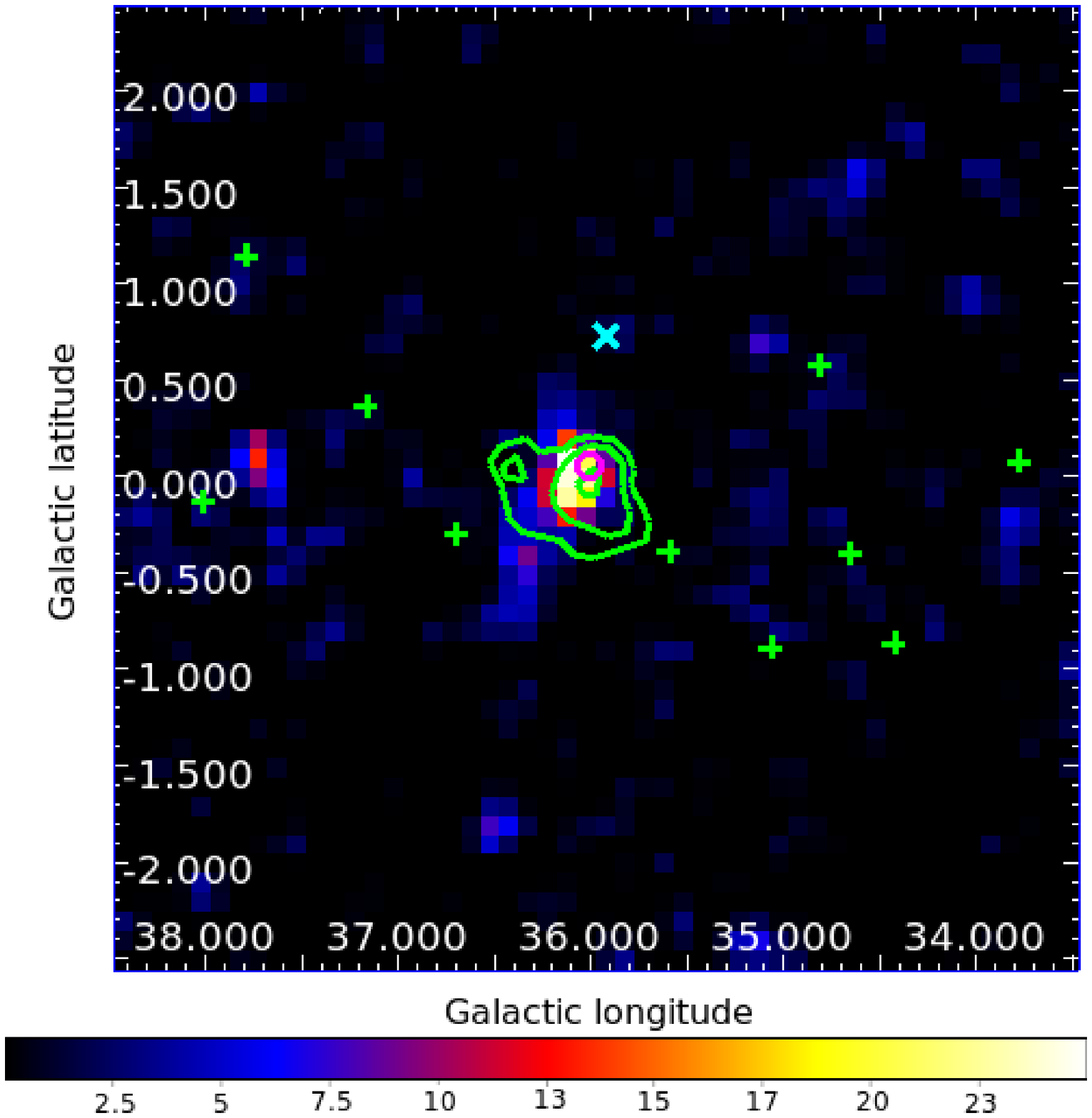}
\label{b}
}

\centering
\subfigure{
\includegraphics[width=0.36\textwidth, height=0.366\textwidth]{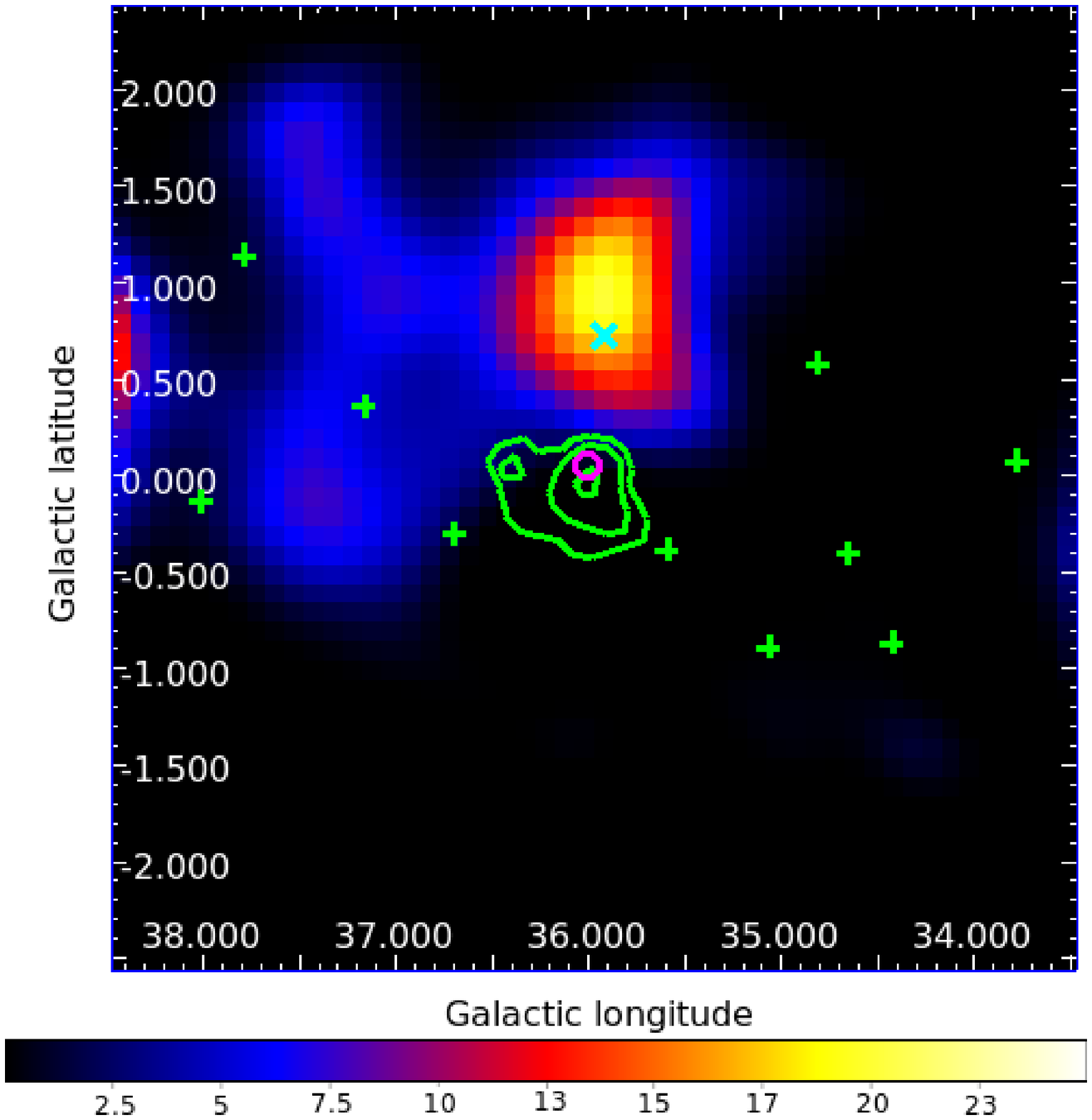}
\label{Bobmap}
}
\centering
\subfigure{
\includegraphics[width=0.36\textwidth, height=0.366\textwidth]{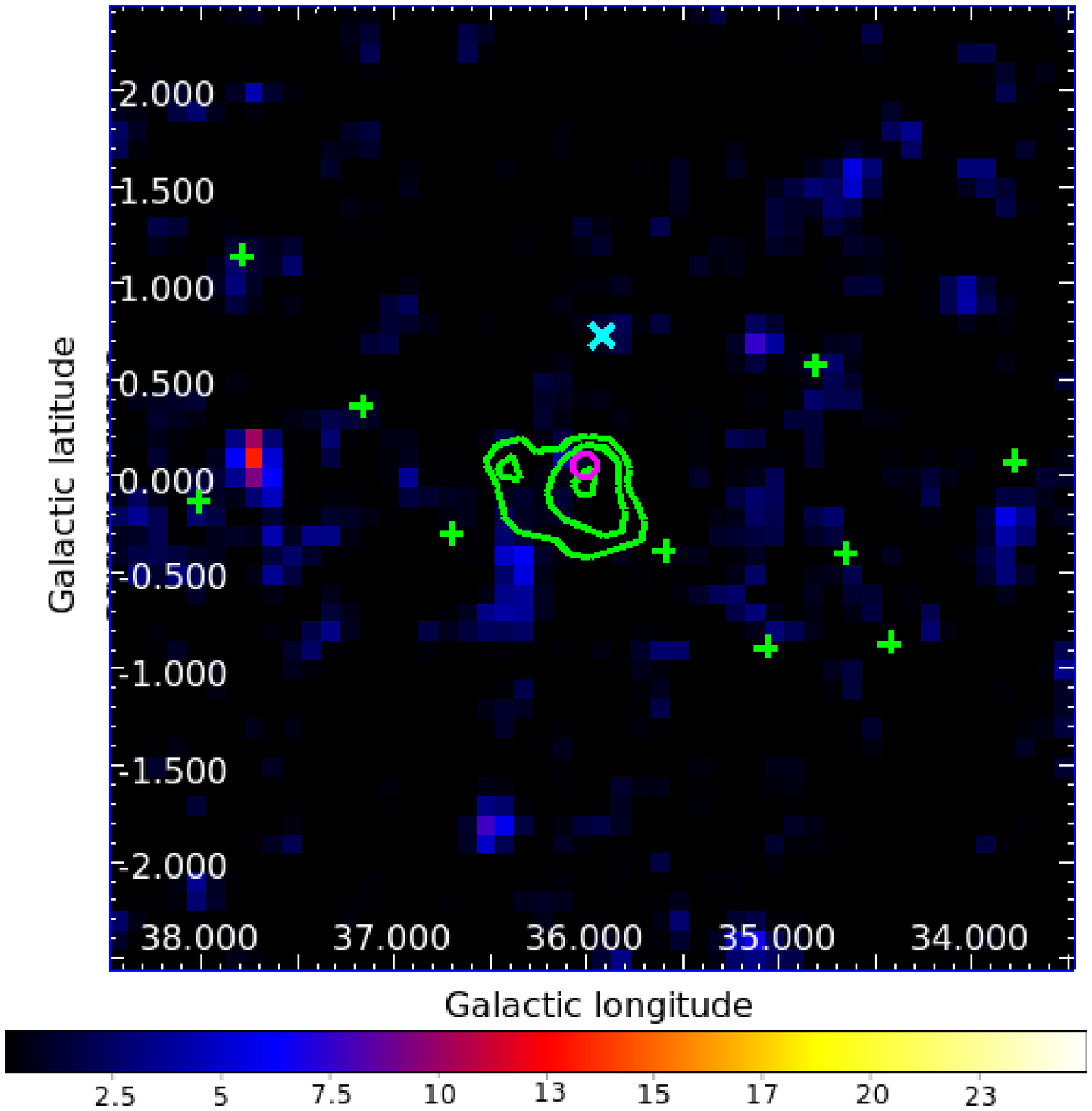}
\label{residual}
}

\caption{TS maps computed by $\mathtt{pointlike}$. The green crosses show the sources of the 2FGL catalog included in the model, whereas the blue X represents the source we added to the model. The green contours show the H.E.S.S. data (Aharonian et al., 2008). The magenta circle gives the position of PSR J1856+0245. \textbf{Top:} TS map obtained between 10 and 300~GeV. The position of the \emph{Fermi}-LAT excess is consistent with that of H.E.S.S. Note that HESS J1857+026 is not included in the model. \textbf{Middle:} TS map obtained between 0.1 and 1.3~GeV, showing the residual excess taken into account in our model. \textbf{Bottom:} Residual TS map obtained between 10 and 300~GeV when all sources are included.}
\end{figure}
\subsection{Spatial and spectral analysis} 

Two different tools were used to perform the spatial and spectral analysis: $\mathtt{gtlike}$ and $\mathtt{pointlike}$. $\mathtt{gtlike}$ is a binned maximum-likelihood method \citep{Mattox 1996} implemented in the Science Tools distributed by the FSSC.  $\mathtt{pointlike}$ is an alternate binned likelihood technique, optimized for characterizing the extension of a source (unlike $\mathtt{gtlike}$), that was extensively tested against $\mathtt{gtlike}$ \citep{Kerr PHD}. These tools fit a source model to the data along with models for the instrumental, extragalactic and Galactic components of the background. In the following analysis, the Galactic diffuse emission is modeled by the standard LAT diffuse emission ring$-$hybrid model \emph{gal\_2yearp7v6\_v0.fits}. The residual cosmic-ray background and extragalactic radiation are described by a single isotropic component with a spectral shape described by the file \emph{iso\_p7v6source.txt}. The models have been released and described by the \emph{Fermi}-LAT Collaboration through the FSSC\footnote{http://fermi.gsfc.nasa.gov/ssc/data/access/lat/BackgroundModels.html}.

The source significance is measured by a test statistic (TS) defined as TS$=2\left(\log \left(L_1\right)-\log \left(L_0\right)\right)$, where $L_1$ corresponds to the likelihood obtained by fitting a model of the source of interest and the background model and $L_0$ corresponds to the likelihood obtained by fitting the background model only. In the following, the  correspondence between the significance and the TS value is evaluated from the $\chi^2$ distribution with 4 degrees of freedom (position and spectral parameters).

The 41 sources within 15$\degr$ of HESS J1857+026 in the Second \emph{Fermi}-LAT catalog \citep{Nolan 2012} were taken into account. We refitted all spectral parameters of the 16 sources within 5$\degr$ around HESS J1857+026. The region includes the bright SNR W44, known to interact with its environment. Extended and only 1.3$\degr$ from HESS J1857+026, W44 could influence our fit. We refitted it assuming an elliptical ring and obtained results consistent with those of \cite{Abdo 2010 a}. The centroid is located at $\alpha=18^{h}56^{m}$ , $\delta=+01\degr22^\prime$. The fitted semi-major and semi-minor axes are respectively maj$/2=(0.33\pm0.10_{stat})\degr$, min/$2=(0.20\pm0.02_{stat})\degr$. The angle of the semi-major axis from celestial North, taken positive toward increasing Right Ascension, is $(327\pm22_{stat})\degr$.

\subsubsection{Shape and position of HESS J1857+026 counterpart}

Source shape analysis requires the best possible angular resolution. Since the source has a hard spectrum (see Section \ref{spec_ana}) we made a compromise between statistics and resolution by selecting photons above 10 GeV. This drastically reduces the contribution of the Galactic diffuse background and improves the single-photon angular resolution. Fig. 1 (Top) presents a LAT TS map in the energy range of 10 GeV to 300 GeV. To each pixel is associated a TS value calculated assuming a point source in its center and fitting only the flux of the source assuming a power-law spectrum with a spectral index of 2. A source coincident with HESS J1857+026 is clearly visible. We determined the extension  of the source using $\mathtt{pointlike}$ with three different models : a point source, a uniform disk and a Gaussian. No significant extension was obtained above 10 GeV. 
The GeV emission was fit to position $\alpha(J2000)=18^{h}57^{m}$ , $\delta(J2000)=+02\degr45^\prime$ with an average statistical error of 0.05$\degr$, consistent with the H.E.S.S. position,  $\alpha(J2000)=18^{h}56^{m}50.80^{s}$ , $\delta(J2000)=+02\degr45^\prime50.2^{\prime \prime}$. 

\subsubsection{Spectral analysis}
\label{spec_ana}

Fig. 1 (Middle) shows a TS map of the region in the energy range 0.1--1.3 GeV. There is excess emission near HESS J1857+026 located at $\alpha(J2000)=18^{h}54^{m}$ , $\delta(J2000)=+02\degr59^\prime$. This excess is inconsistent with that of HESS J1857+026 and was added to the background model. 
This additional background source was fitted assuming a pure power-law with an integrated flux of F(0.1--100 GeV)=$(2.31 \pm 0.37_{\rm stat})\times 10 ^{-7}$ ph cm$^{-2}$ s$^{-1}$, a spectral index of $\Gamma = 3.18 \pm 0.56_{\rm stat}$, which gives a significance above 300 MeV of $\sim 3.6 \sigma$ (TS=20).

Spectral analysis was performed using $\mathtt{gtlike}$, selecting only 0.3--300 GeV to avoid the low energy range that is dominated by the diffuse Galactic background and subject to large systematic uncertainties. In this energy range, HESS J1857+026 is well described by a pure power-law with an integrated flux extrapolated down to 100 MeV of F(0.1--100 GeV)=$(5.78 \pm 0.85_{\rm stat} \pm 3.11_{\rm syst})\times 10 ^{-9}$ ph cm$^{-2}$ s$^{-1}$, a spectral index of $\Gamma = 1.53 \pm 0.11_{\rm stat} \pm 0.55_{\rm syst}$ and an energy flux of G(0.1--100 GeV)=$(2.71 \pm 0.52_{\rm stat} \pm 1.51_{\rm syst})\times 10 ^{-11}$ ergs cm$^{-2}$ s$^{-1}$, which gives a significance above 300 MeV of $\sim 5.4 \sigma$ (TS=39). The residual TS map after fit in Fig.1 (Bottom) shows no significant excess. 

\emph{Fermi}-LAT spectral points for HESS J1857+026 were obtained by splitting the 0.3--100 GeV range into 4 logarithmically-spaced energy bins plus a bin between 100 and 300 GeV which contains 22 photons corresponding to a TS of 14, shown in Fig. 2. A 99~\% C.L. upper limit is computed when TS$<$10 using the approach of \cite{Nolan 2012}. The errors on the spectral points represent the statistical and systematic uncertainties added in quadrature. 

Three main systematic uncertainties can affect the LAT flux estimate for a point source: uncertainties in the Galactic diffuse background, uncertainties on the effective area and uncertainties on the source shape. The dominant uncertainty at low energy comes from the Galactic diffuse emission, estimated by  changing the normalization of the Galactic diffuse model artificially by $\pm6\%$ as in (Abdo 2010 b). Since it is computed for P6 IRFs, the 6\% factor overestimates the bias. The second systematic is estimated by using modified IRFs. The fact that we do not know the true $\gamma$-ray morphology introduces a last source of error. We derived an estimate of the uncertainty on the source shape by using the best Gaussian model obtained by H.E.S.S. We combine these various errors in quadrature to obtain our best estimate of the total systematic error at each energy, which we propagate through to the fit model parameters.

Assuming a distance of 9~kpc and isotropic emission, the $\gamma$-ray flux corresponds to $L_{PWN}^{\gamma} (0.1$--$100$ GeV$)=(2.5 \pm 0.5_{stat} \pm 1.5_{syst}) \times 10^{35} \left( \frac{d}{9 kpc} \right)^2$ ergs s$^{-1}$. Using the pulsar's $\dot{E}$, yields a $\gamma$-ray efficiency of $\sim$ 5 \%, one of the highest PWN efficiencies observed at GeV energies \citep{Ackermann 2011}. It is still in the range of expected values for PWNe seen by the \emph{Fermi}-LAT and is close to the estimate of 3\% using H.E.S.S. data \citep{Mattana09}.

\section{Supporting X-Ray measurement}
To find the flux of a potential X-ray PWN associated with PSR J1856+0245, we analyzed a 39-ks {\it Chandra} ACIS-I observation from February 28, 2011 (Obs. ID 12557).  These data were recorded in the VFAINT and Timed Exposure (TE) modes and were analyzed using CIAO\footnote{Chandra Interactive Analysis of Observations \citep{2006SPIE6270E60F}.}  version 4.3.1 with CALDB 4.4.3. PSR J1856+0245 is clearly detected as a point source, but there was no immediate evidence for extended emission surrounding this position. 
Based on a 30-ks XMM-Newton EPIC pn observation (ObsID 0505920101), the unabsorbed flux of the pulsar in the 2--10 keV range, assuming a power-law spectrum, is $8.3^{+2.5}_{-7.9} \times 10^{-14}$ ergs s$^{-1}$. A more in-depth analysis of the X-ray properties of PSR J1856+0245 as well as the XMM data will be presented in \cite{Bogdanov}.

Given that the size of the potential X-ray PWN is not known, we investigated an appropriately sized extraction region to see whether it contains statistically significant count excess above background. The extraction region is an annulus extending from $2^{\prime \prime}$ -- $15^{\prime \prime}$ from the position of the pulsar. The inner radius was chosen to avoid contamination from the pulsar. The outer radius was chosen based on the X-ray PWNe observed for pulsars with comparable $\dot{E}$ by scaling their angular size with their distance \citep{Kargaltsev 2008}.  The background regions were chosen from several other source-free regions near the pulsar. For the $2^{\prime \prime}$ -- $15^{\prime \prime}$ extraction region we find an upper limit on the unabsorbed flux of $5 \times 10^{-14}$\,erg s$^{-1}$ cm$^{-2}$ ($1$--$10$\,keV, 3$\sigma$ confidence), corresponding to a luminosity of $5 \times 10^{32}$\,erg s$^{-1}$.  The counts in this region show a 2$\sigma$ excess from zero counts.  Assuming instead the TeV position and a $6^\prime$ extraction region consistent with the H.E.S.S. morphology, we find an upper limit on the unabsorbed flux of $2 \times 10^{-12}$\,erg s$^{-1}$ cm$^{-2}$ ($1$--$10$\,keV, 3$\sigma$ confidence), corresponding to a luminosity of $2 \times 10^{34}$\,erg s$^{-1}$. Given the marginal significance of the count excesses derived, we cannot convincingly claim the detection of a weak X-ray PWN. These luminosity limits are derived from the 3$\sigma$ upper bound on the net count rate and assume a typical power-law spectrum of index 1.5 for the PWN, a distance of 9 kpc, and a column density $N_{\rm H} = 4 \times 10^{22}$\,cm$^{-2}$ based on the spectroscopic analysis of the XMM data. 
\begin{figure}[t!]

\subfigure{
\includegraphics[width=0.5\textwidth]{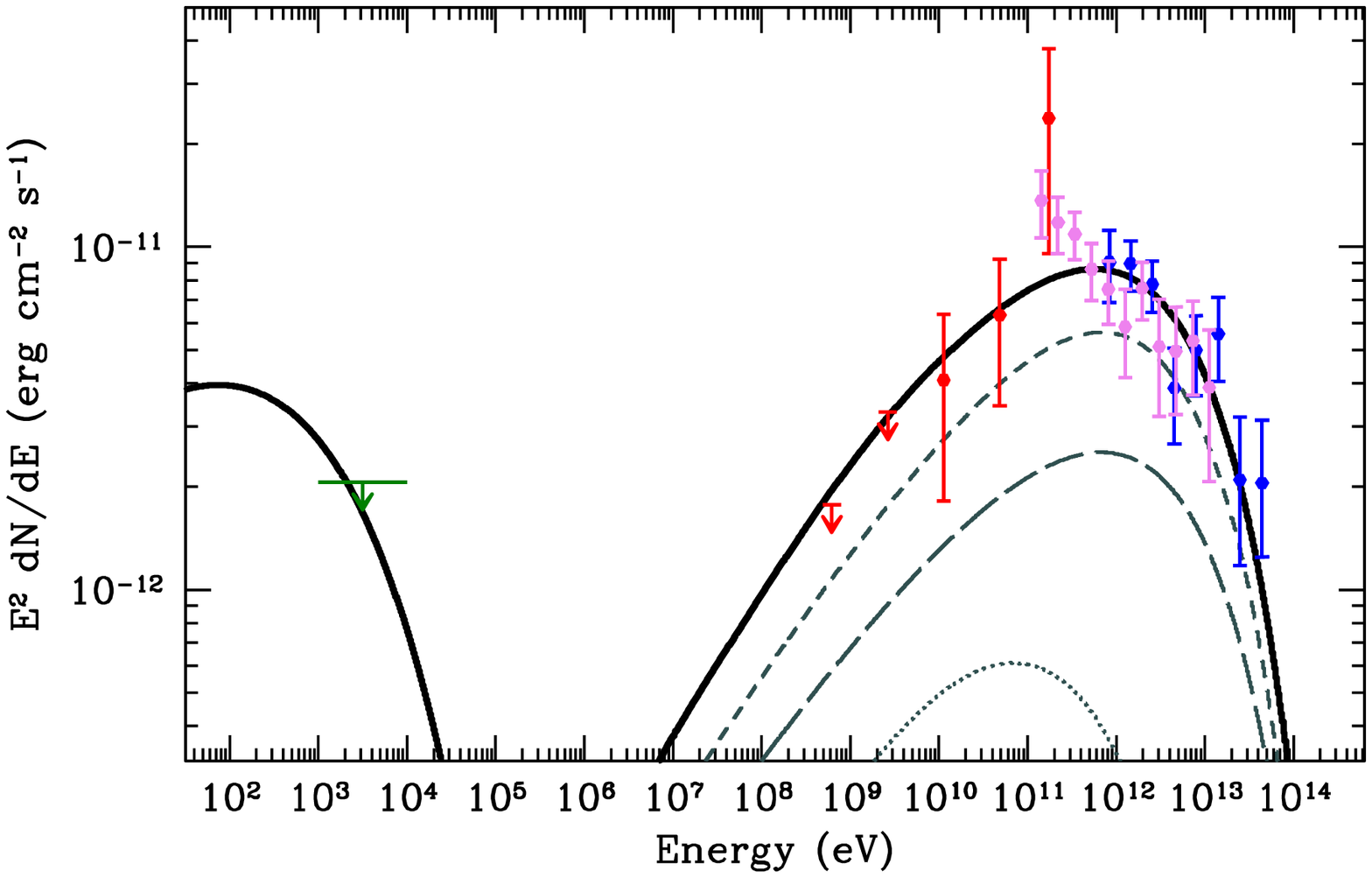}
\label{fig:powmodel}
}

\subfigure{
\includegraphics[width=0.5\textwidth]{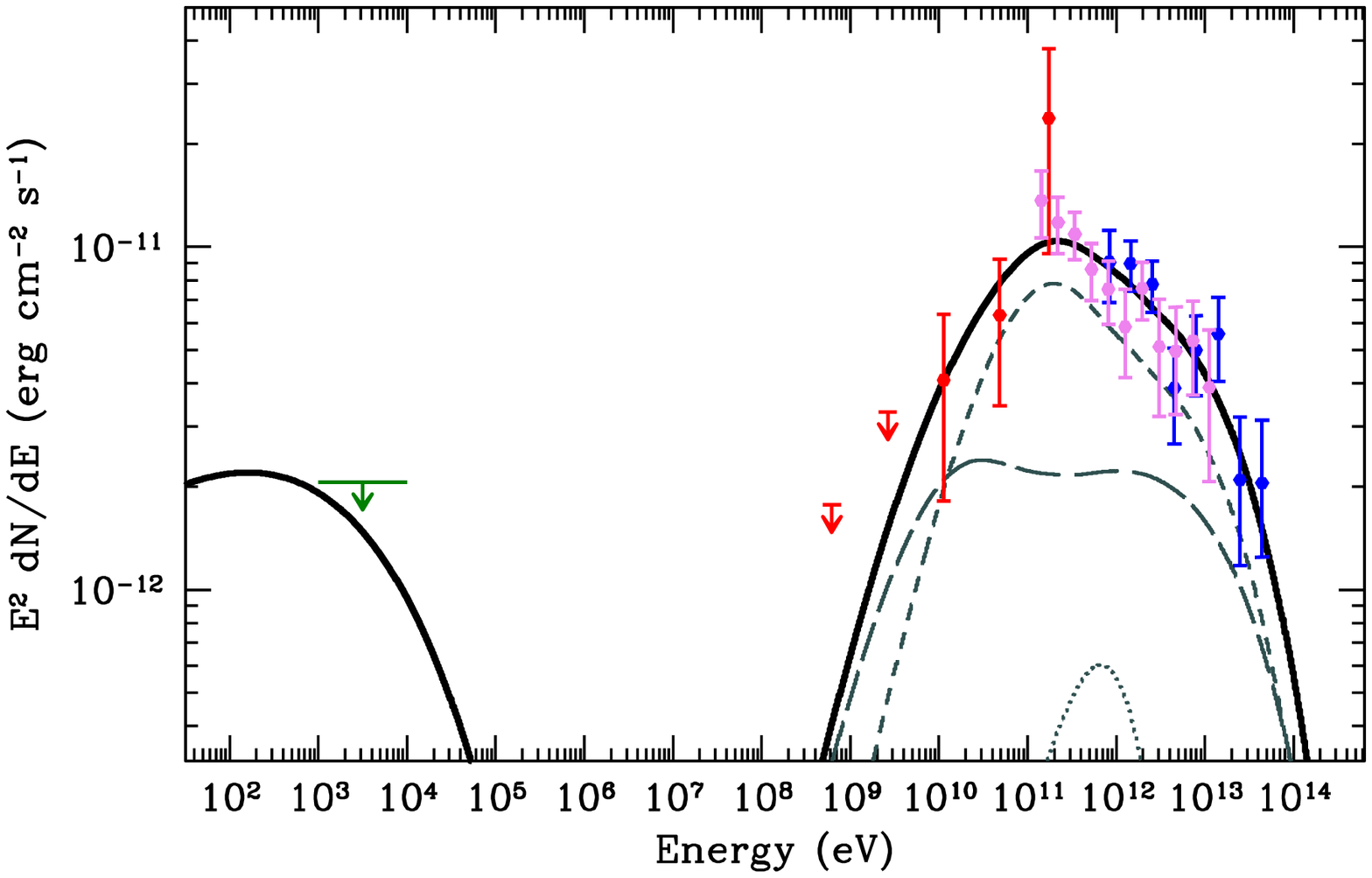}
\label{fig:maxmodel}
}

\subfigure{
\includegraphics[width=0.5\textwidth]{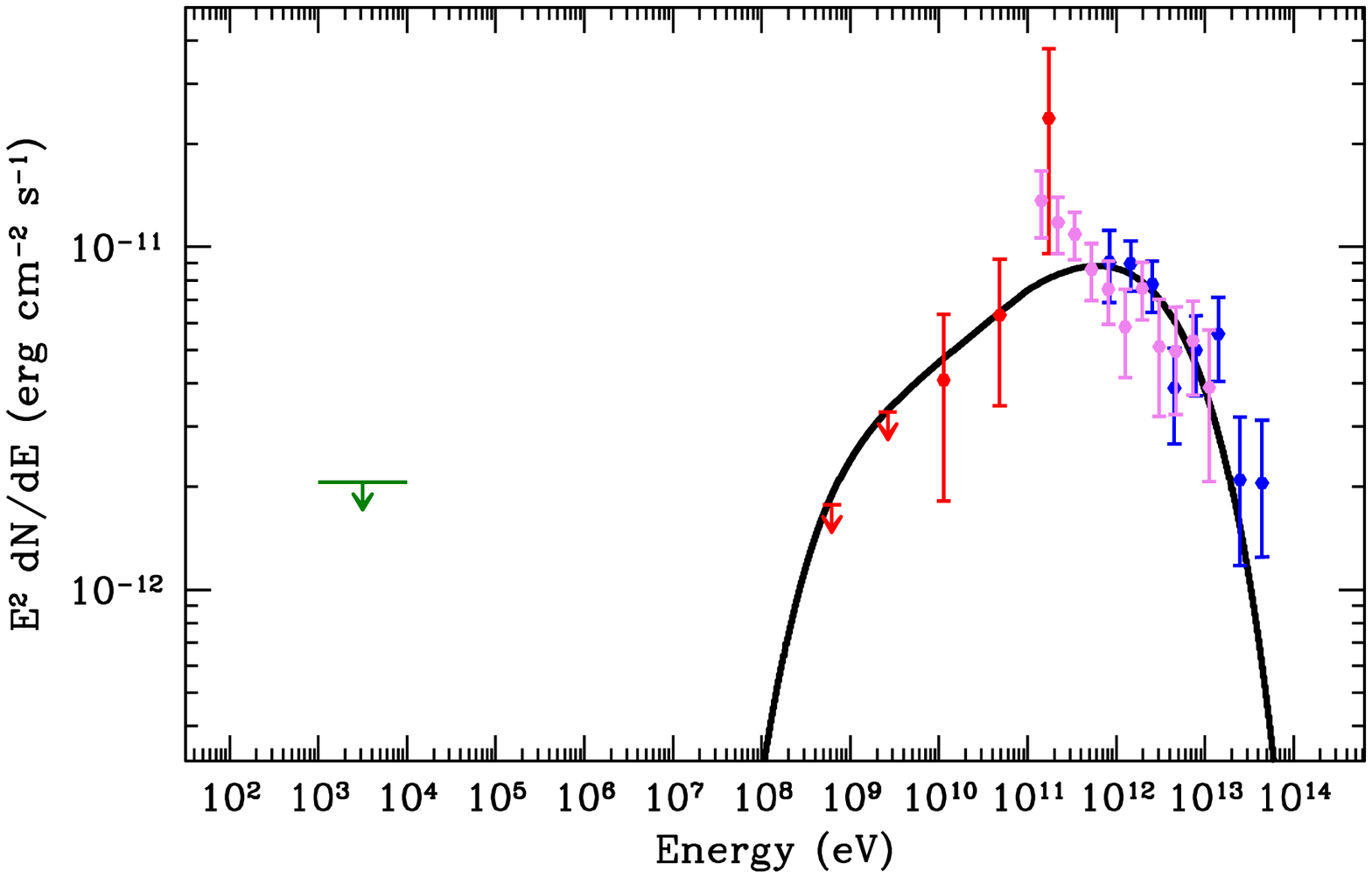}
\label{fig:hadronmodel}
}
\caption{\label{fig:powmodel}Spectral energy distribution of
HESS~J1857$+$026 with a simple exponentially cutoff power-law electron spectrum \textbf{(Top)}, a relativistic Maxwellian plus power-law electron spectrum \textbf{(Middle)}, and an exponentially cutoff power-law proton spectrum \textbf{(Bottom)}. The X-ray flux upper limit obtained using Chandra(green), LAT spectral points (red), MAGIC points (violet) \citep{Klepser 2011}, and H.E.S.S. points (blue) \citep{Aharonian 2008} are shown. The black line denotes the total synchrotron, inverse Compton and pion decay emission from the nebula.Thin curves indicate the Compton components from scattering on the CMB (long-dashed), IR (medium-dashed), and stellar (dotted) photons.}
\end{figure}

\section{Discussion}

The nature of HESS~J1857+026 remains unclear. From the observations presented here, we know that the GeV source is positionally and spectrally consistent with the TeV source, suggesting a physical relationship. The limits on an X-ray PWN indicate a low magnetic field for any leptonic model, because a larger field would produce strong X-ray synchrotron emission. The spatial extent of the TeV source and the proximity to a pulsar suggests a PWN. The distance, characteristic age, and energy available from the pulsar are known, although any distance estimate based on DM has a significant uncertainty. Using these inputs, we investigate whether a plausible PWN model consistent with all the observations can be found.

To investigate the global properties of the PWN, we apply a one-zone time dependent Spectral Energy Distribution (SED) model which reproduces the multi-wavelength measurements from MAGIC \citep{Klepser 2011}, H.E.S.S. \citep{Aharonian 2008}, as well as the LAT results and the $2^{\prime \prime}$ -- $15^{\prime \prime}$ X-ray upper limit described above. This model is described in \cite{grondinetal11} and \citet{velax}, with a more detailed description (applied to a multi-zone scenario rather than one-zone) found in Section 3.10 of \citet{Van Etten 2011}.

The model computes SEDs from evolving electron populations over the lifetime of the pulsar in a series of time steps with the energy content of the injected particle population varying with time according to the pulsar spin down. At each time step, new particles are injected and synchrotron, inverse Compton, and adiabatic cooling is computed. Synchrotron and inverse Compton fluxes are computed from the final electron spectrum, magnetic field, and photon fields. Starting assumptions include: the source of all particles and magnetic field in the nebula is PSR~J1856+0245, ambient photon fields are uniform and do not vary over the evolution time of the electron populations, and the nebula distance is 9 kpc.

The large size of the nebula ($\sim 20$ pc at 9 kpc) implies a middle-aged PWN, whose parent SNR has likely evolved from the free-expansion phase to the Sedov phase.  The Sedov phase is expected to occur on a timescale of $t_{Sedov}\approx3$ kyr for an explosion of $10^{51}$ erg, an ejecta mass of $10 M \odot$, and an ambient medium density of $0.1 \, \rm{cm^{-3}}$ \citep{tm99}. Eventually, the inward moving SNR reverse shock collides with the expanding PWN, which can happen as late as 5 times the transition to the Sedov phase \citep{vanderswaluw04}.  
The interaction of the PWN and the SNR reverse shock compresses the PWN, resulting in an increased magnetic field. Even in the spherically symmetric case the evolution is complex, with the PWN undergoing a series of oscillations due to the reverse shock interaction. The significant offset of the pulsar from the $\gamma$-ray centroid implies either a PWN expansion into an inhomogenous medium, or an asymmetric reverse shock interaction.  Either way, models such as \citet{gelfandetal09} which compute the PWN compression assuming spherical symmetry cannot be applied.

Given the lack of multi-wavelength data for HESS J1857+026 and the complexity of the SNR-PWN interaction, we see little reason to adopt a complicated spatial model and therefore assume a simple spatial evolution. During the free-expansion phase of the PWN (assumed to be $\sim 3$ kyr) we adopt an expansion of $R \propto t$, following which the radius evolves as $R \propto t^{0.3}$, appropriate for a PWN expanding in pressure equilibrium with a Sedov phase SNR \citep{rc84}.
We also adopt a simple magnetic field model, and assume that over the pulsar lifetime the magnetic field evolves as $B \propto t^{-1.5}$, following $\sim$ 500 years of constancy. This magnetic field evolution is similar to the $t^{-1.3} - t^{-2}$ behavior derived by \citet{rc84} and the $t^{-1.7}$ evolution of \citet{gelfandetal09} during the initial expansion phase. The power-law evolution of the magnetic field is similar to that adopted by \citet{zhang08}, \citet{lem09}, \citet{mayer12} for similarly aged PWNe.

Model fitting is achieved by minimizing the $\chi^{2}$ between model and data using the downhill simplex method described in \citet{pressetal92}. For each ensemble of $N$ variable parameters we evolve the system over the pulsar lifetime and calculate $\chi^{2}$ between model curves and flux data points. The simplex routine subsequently varies the parameters of interest
to minimize the fit statistic. We estimate parameter errors by computing $\chi^{2}$ for a sampling of points near the best fit values and using these points to fit the $N-$dimensional 
ellipsoid describing the surface of $\Delta \chi^{2} = 2.71$. Assuming that the adopted PWN model is correct, and that errors are Gaussian, the projected size of this $\Delta \chi^{2} = 2.71$ ellipsoid onto each parameter axis defines the 90\% multi-parameter (projected) error. 

Using the radio pulsar ephemeris, we tried to derive a value for the braking index using Eq. \ref{bracking} where $\nu$, $\nu^{(i)}$ and n represent respectively the rotational frequency of the pulsar, its time derivative of order i and the braking index.

\begin{equation}
\label{bracking}
n=\frac{\nu \nu^{(2)}}{\left(\nu^{(1)}\right)^2}
\end{equation}

We obtained n$>$20. Large braking indices between glitches are common among Vela-like pulsars and are likely to be associated with glitch recoveries. These large values should not be interpreted as the long-term braking index due to secular spin evolution but instead these correspond to transient states caused by large glitch activity as discussed in section 3.2.2 of \cite{Hobbs 2010}. Dipole braking indices have been measured only for a few pulsars with the highest spindown rates (see Table 1 of \cite{espinoza11}).

With its characteristic age of $\sim$21 kyr, PSR~J1856+0245 is a Vela-like pulsar possibly affected by  large glitch activity, which we have not yet directly seen. Thus, we fix here the pulsar braking index to the canonical value of 3.

We assume the existence of three primary photon fields (CMBR, far IR (dust), and starlight) and use the interstellar radiation field from GALPROP\footnotetext{http://galprop.stanford.edu/resources.php?option=data} \citep{porteretal05} to estimate the photon fields at the Galactic radius of PSR\, J1856$+$0245. A distance of 9~kpc in the direction of the pulsar corresponds to a Galactic radius of 5.4~kpc. At this radius, the peak of the SED of dust IR photons corresponds to a black-body temperature of $T \sim 32$ K with a density of $\sim 1.1$ eV$\rm \, cm^{-3}$, while the SED of stellar photons peaks at $T \sim 2500$ K with a density of $\sim 1.2$ eV$\rm \, cm^{-3}$.

A simple exponentially cutoff power-law injection of electrons, evolved properly over the pulsar lifetime, often provides an adequate match to PWN SEDs. For this injection spectrum we fit four variables: final magnetic field $B_{\rm{f}}$, electron high energy cutoff $E_{\rm{cut}}$, electron power-law index $p$, and initial pulsar spin period $P_0$.  The best fit parameters, with errors, are given in Table 2.  The initial spin period gives an age of $20\pm 2$ kyr. This model poorly matches the low energy MAGIC points,
as shown in Figure~\ref{fig:powmodel} (Top). The low value of $\sim$ 4 $\mu$G found for the magnetic field is consistent with 6 $\mu$G found for the 15 kyr old PWN HESS J1640--465 in \citet{lem09}

Another option to fit the multi-wavelength data is to adopt the relativistic Maxwellian plus power-law tail electron spectrum proposed by \cite{spitkovsky08}.  We implement this spectrum as described in \cite{grondinetal11}. 
The best fit, presented in Fig.~\ref{fig:powmodel} (Middle) and Table 2, is obtained with $kT = 0.64$ TeV corresponding to an upstream Lorentz factor of $2.5 \times 10^{6}$.  A very high cutoff of 390 TeV is required, though the power-law index of $p=2.44$ is consistent with the value of $\sim 2.5$ proposed by \cite{spitkovsky08}. The initial spin period gives an age of $15\pm2$ kyr. The relativistic Maxwellian plus power-law model better matches the multi-wavelength data, and also directly probes the upstream pulsar wind via fitting of the upstream Lorentz factor of the wind.

A hadronic scenario is also possible, with $\gamma$-rays arising from proton-proton interactions. 
For this model, corresponding to Fig.~\ref{fig:powmodel} (bottom) and Table 3, we fix the ambient gas density at $50 \, \rm{cm}^{-3}$ and age at 20 kyr. It should be noted though that the total energy injected in such scenario is very high even for the large density assumed here.

\begin{table*}[h!]

\caption{Model Parameters : the leptonic fits, including the statistical uncertainties on the fitted parameters.}
\begin{tabular}{cccccccccc}
\hline 
\hline
Model $\T \B$ & $n ^{a}$ & $t_{Sedov}^{b}$ $\, (\rm{kyr})$ & $\beta \, (B \propto r^{\beta})$ & $B_f \, (\mu\rm{G})$ & $p \, (E^{-p})$ &  $kT \, (\rm{TeV})$ & $E_{\rm{cut}}$ (TeV) & $P_0 \, (\rm{ms})$ & $\chi^2$/d.o.f. \\
\hline
Power-Law \T & $3^{c}$ & $3^{c}$ & $-1.5^{c}$ & $3.9 \pm 0.4$ & $2.12 \pm 0.03$ & - & $120 \pm 44$ & $11.1 \pm 6.8$ & $21.9/21$ \\
Rel. Max. & $3^{c}$ & $3^{c}$ & $-1.5^{c}$ & $2.9 \pm 0.6$ & $2.44 \pm 0.10$ & $0.64 \pm 0.09$ & $390 \pm 280$ & $42 \pm 3$ & $12.5/20$ \\
\hline
\end{tabular}\newline 
$^{a}$ Pulsar braking index\newline
$^{b}$ Sedov phase onset, after which $r\propto t^{0.3}$\newline
$^{c}$ Held fixed
\label{table2}

\end{table*}

\begin{table*}[h!]

\caption{Model Parameters : the hadronic fit, including the statistical uncertainties on the fitted parameters.}
\begin{tabular}{cccccccccc}
\hline 
\hline
Model $\T \B$ & $n ^{a}$ & $t_{Sedov}^{b}$ $\, (\rm{kyr})$ & $\beta \, (B \propto r^{\beta})$ & $B_f \, (\mu\rm{G})$ & $p \, (E^{-p})$ &  $n \, (\rm{cm}^{-3})^{c}$ & $E_{\rm{cut}}$ (TeV) & $E_0 (\times 10^{50} \rm{erg}) ^{d}$ & $\chi^2$/d.o.f. \\
\hline
Hadron \B & $3^{e}$ & $3^{e}$ & $-1.5^{e}$ & $20_{-20}^{+80}$ & $1.83 \pm 0.04$ & $50^{c,e}$ & $75 \pm 25$ & $ 0.64 \pm 0.06^{d}$ & $24.6/21$ \\
\hline
\end{tabular}\newline 
$^{a}$ Pulsar braking index\newline
$^{b}$ Sedov phase onset, after which $r\propto t^{0.3}$\newline
$^{c}$ Ambient medium density\newline
$^{d}$ Total energy injected\newline
$^{e}$ Held fixed
\label{table3}

\end{table*}
\normalsize

\section{Conclusions}
Using 3 years of \emph{Fermi}-LAT data, a $\gamma$-ray source has been detected at high significance at a position coincident with the TeV source HESS J1857+026. We have investigated whether a model in which a PWN is powered by the pulsar PSR~J1856+0245 can reproduce the multi-wavelength data. In such a scenario, the VHE spectrum observed by MAGIC and H.E.S.S., combined with the limits imposed by the steep LAT data, is difficult to match with a simple power-law injection of electrons (or protons), and we find a significantly better fit with a relativistic Maxwellian plus power-law spectrum. The low magnetic field of these leptonic fits, due to the stringent X-ray upper limit, implies that if PWN leptons are indeed responsible for the $\gamma$-ray flux, they must be dominated by relic electrons which have escaped the PWN core into weakly magnetized surroundings. The hadronic scenario relaxes this constraint, though the energy requirements are quite high even for a dense ambient medium, and a very hard power-law index is required. At present the true nature of HESS J1857+026 remains a mystery, though the new LAT data and X-ray upper limit hint that this source may be another relic PWN, increasing the population of such high energy $\gamma$-ray systems.

\emph{Acknowledgements}  The \textit{Fermi}-LAT Collaboration acknowledges generous ongoing support
from a number of agencies and institutes that have supported both the development and the operation of the LAT as well as scientific data analysis. These include the National Aeronautics and Space Administration and the Department of Energy in the United States, the Commissariat \`a l'Energie Atomique and the Centre National de la Recherche Scientifique / Institut National de Physique Nucl\'eaire et de Physique des Particules in France, the Agenzia Spaziale Italiana and the Istituto Nazionale di Fisica Nucleare in Italy, the Ministry of Education, Culture, Sports, Science and Technology (MEXT), High Energy Accelerator Research Organization (KEK) and Japan Aerospace Exploration Agency (JAXA) in Japan, and the K.~A.~Wallenberg Foundation, the Swedish Research Council and the Swedish National Space Board in Sweden.

Additional support for science analysis during the operations phase is gratefully acknowledged from the Istituto Nazionale di Astrofisica in Italy and the Centre National d'\'Etudes Spatiales in France.

The Lovell Telescope is owned and operated by the University of Manchester as part of the Jodrell Bank Centre for Astrophysics with support from the Science and Technology Facilities Council of the United Kingdom.

\bibliographystyle{aa}

\begin{thebibliography}{}
  
  \bibitem[{Abdo et al., 2009a}]{Abdo 2009 a}
  Abdo, A. A., Ackermann, M., Ajello, M., et al.
  \newblock 2009a, \apj, 696, 1084  
  
  \bibitem[{Abdo et al., 2009b}]{Abdo 2009 b}
  Abdo, A. A., Ackermann, M., Ajello, M., et al.
  \newblock 2009b, Phys. Rev. D, 80, 122004 
  
  \bibitem[{Abdo et al., 2010a}]{Abdo 2010 a}
  Abdo, A. A., Ackermann, M., Ajello, M., et al.
  \newblock 2010a, Science, 327, 1103 
  
  \bibitem[{Abdo et al., 2010b}]{Abdo 2010 b}
  Abdo, A. A., Ackermann, M., Ajello, M., et al.
  \newblock 2010b, \apj, 722, 1303 
  
  \bibitem[{Abdo et al., 2010c}]{Abdo 2010 c}
  Abdo, A. A., Ackermann, M., Ajello, M., et al.
  \newblock 2010c, \apj S, 187, 460 

  \bibitem[Abdo et al., 2010]{velax} 
  Abdo, A. A., Ackermann, M., Ajello, M., et al.
  \newblock 2010d, ApJ, 713, 146
   
  \bibitem[Ackermann et al., 2011]{Ackermann 2011}
  Ackermann, M., Ajello, M., Baldini, L., et al.
  \newblock 2011, \apj, 726, 35 

  \bibitem[Aharonian et al., 2008]{Aharonian 2008}
  Aharonian, F., Akhperjanian, A. G., Barres de Almeida, U., et al. 
  \newblock 2008, A\&A, 477, 353 
  
  \bibitem[Atwood et al., 2009]{Atwood 2009}
  Atwood, W. B., Abdo, A. A., Ackermann, M., et al.
  \newblock  2009, \apj, 697, 2, 1071
  
  \bibitem[Bogdanov et al., in preparation]{Bogdanov}
  Bogdanov, S., et al.
  \newblock in preparation 
  
  \bibitem[Cordes et al., 2002]{Cordes 2002}
  Cordes, J. M., \& Lazio, T. J. W.
  \newblock 2002, arXiv:astro-ph/0207156
  
  
  \bibitem[Espinoza et al., 2011]{espinoza11}
  Espinoza, C. M., Lyne, A. G., Kramer, M., Manchester, R. N., \& Kaspi V. M.
  \newblock 2011, \apj L, 741, L13 
  
  \bibitem[Fruscione et al.(2006)]{2006SPIE6270E60F} 
  Fruscione, A., McDowell, J. C., Allen, G. E., et al.
  \newblock 2006, \procspie, 6270 
  
  \bibitem[Gaensler el al., 2006]{Gaensler 2006}
  Gaensler, B. M., \& Slane, P.~O.
  \newblock 2006, Ann. Rev. of A\&A, 44, 17
  
  \bibitem[Gelfand et al.(2009)]{gelfandetal09} 
  Gelfand, J.~D., Slane, P.~O., \& Zhang, W.
  \newblock 2009, \apj, 703, 2051  
  
  \bibitem[Grondin et al., 2011]{grondinetal11} 
  Grondin, M.-H., Funk, S., Lemoine-Goumard, M., et al. 
  \newblock 2011, \apj, 738, 42  
  
  \bibitem[Hinton et al., 2010]{Hinton 2010}
  Hinton, J.A., \& Hofmann, W. 
  \newblock 2010, Ann. Rev. A\&A, 47, 523
  
  \bibitem[Hobbs et al., 2004]{Hobbs 2004}
  Hobbs, G., Lyne, A. G., Kramer, M., Martin, C. E., \& Jordan, C.
  \newblock 2004, MNRAS, 353, 1311 
  
  \bibitem[Hobbs et al., 2006]{Hobbs 2006}
  Hobbs, G., Edwards, R., \& Manchester, R.
  \newblock 2006, Chin. J. A A, 6, 2, 189
  
  \bibitem[Hobbs et al., 2010]{Hobbs 2010}
  Hobbs, G., Lyne, A. G., \& Kramer, M.
  \newblock 2010, MNRAS, 402, 1022
  
  
  \bibitem[Hessels et al., 2008]{Hessels 2008}
  Hessels J. W. T., Nice, D. J., Gaensler, B. M., et al.
  \newblock 2008, \apj, 682, L41
  
  \bibitem[Kerr, 2011]{Kerr PHD}
  Kerr, M.
  \newblock 2011, arXiv:1101.6072v1
  
  \bibitem[Klepser et al., 2011]{Klepser 2011}
  Klepser, S., Krause, J., \& Michele, D.
  \newblock 2011, proceedings of the 32th  International Cosmic Ray Conference : arXiv:1109.6448v1

  \bibitem[Kargaltsev \& Pavlov, 2008]{Kargaltsev 2008}
  Kargaltsev, O. \& G. G. Pavlov, G. G.
  \newblock 2008, arXiv:0801.2602v2
  
  \bibitem[Lemiere et al.(2009)]{lem09} 
  Lemiere, A., Slane, P.~O., Gaensler, B.~M., \& Murray, S.
  \newblock 2009, \apj, 706, 1269
  
  \bibitem[{Mattana et al., 2009}]{Mattana09}
  Mattana, F., Falanga, M., Götz, D., et al.
  \newblock 2009, \apj, 694, 12   
  
  \bibitem[{Mattox et al., 1996}]{Mattox 1996}
  Mattox, J. R., Bertsch, D. L., Chiang, J. et al.
  \newblock 1996, \apj, 461, 396
  
  \bibitem[Mayer et al.(2012)]{mayer12} 
  Mayer, M., Brucker, J., Jung, I., Valerius, K., \& Stegmann, C.
  \newblock 2012, arXiv:1202.1455
  
  \bibitem[Neronov et al., 2010]{Neronov 2010}
  Neronov A., \& Semikoz, D.
  \newblock 2010, arXiv:1011.0210v1
  
  \bibitem[{Nolan et al., 2012}]{Nolan 2012}
  Nolan, P. L., Abdo, A. A., Ackermann, M., et al. 
  \newblock 2012, \apj S, 199, 31
  
  \bibitem[Porter et al., 2005]{porteretal05}   	
  Porter, T. A., \& Strong, A. W.
  \newblock 2005, proceedings of the 29th International Cosmic Ray Conference, 4, 77
  
  \bibitem[Press et al.(1992)]{pressetal92} 
  Press, W.~H., Teukolsky, S.~A., Vetterling, W.~T., \& Flannery, B.~P. 
  \newblock 1992, Cambridge: University Press, 2nd ed.
  
  \bibitem[Reynolds \& Chevalier(1984)]{rc84} 
  Reynolds, S.~P., \& Chevalier, R.~A.
  \newblock 1984, \apj, 278, 630
  
  \bibitem[Romani et al., 2011]{Romani11} 
  Romani, R. W., Kerr, M., Craig, H. A., et al. 
  \newblock 2011, \apj, 738, 114 
  
  \bibitem[Slane et al., 2010]{slane10} 
  Slane, P.~O., Castro, D., Funk, S., et al. 
  \newblock 2010, \apj, 720, 266
  
  \bibitem[Smith et al., 2008]{Smith 2008}
  Smith, D. A., Guillemot, L., Camilo, F., et al.
  \newblock 2008, A\&A, 492, 923  

  \bibitem[Spitkovsky, 2008]{spitkovsky08} 
  Spitkovsky, A.
  \newblock 2008, \apj L, 682, L5 
  
  \bibitem[Tanaka \& Takahara(2011)]{tt11} 
  Tanaka, S.~J., \& Takahara, F.
  \newblock 2011, \apj, 741, 40
  
  \bibitem[{Torres et al., 2011}]{Torres 2011}
  Torres, D. F., Li, H., Chen, Y., et al. 
  \newblock 2011, MNRAS, 417, 3072
  
  \bibitem[Truelove \& McKee(1999)]{tm99} 
  Truelove, J.~K., \& McKee, C.~F.\
  \newblock 1999, \apjs, 120, 299
  
  \bibitem[van der Swaluw et al.(2004)]{vanderswaluw04} 
  Van der Swaluw, E., Downes, T.~P., \& Keegan, R.
  \newblock 2004, \aap, 420, 937 
  
  

  
 \bibitem[{Van Etten  \& Romani(2011)}]{Van Etten 2011}
 Van Etten, A. \& Romani, R. W.
 \newblock 2011, ApJ, 742, 62
  
 \bibitem[Zhang et al.(2008)]{zhang08} 
 Zhang, L., Chen, S.~B., \& Fang, J.
 \newblock 2008, \apj, 676, 1210



\end{thebibliography}
 
\end{document}